# Planetary Defense Use of the SPHEREx Solar System Object Catalog

Carey M. Lisse, Space Exploration Sector, Johns Hopkins University Applied Physics Laboratory, Laurel, MD, 20723 USA

James Bauer, Yaeji Kim, Department of Astronomy, University of Maryland at College Park, College Park, MD 20207 USA

**Abstract:** The upcoming 2025 – 2027 NASA SPHEREx (Spectro-Photometer for the History of the Universe, Epoch of Reionization, and Ices Explorer) MIDEX all-sky 0.7 – 5.0 um spectral survey provides a unique space-based opportunity to detect, spectrally categorize, and catalog hundreds of thousands of solar system objects at NEOWISE sensitivities. This paper discusses the unique near-infrared capabilities of SPHEREx, its potential applications in Planetary Defense, (PD), and the implications for risk mitigation associated with Potentially Hazardous Objects (PHOs). By leveraging SPHEREx data, scientists and decision-makers can enhance our ability to track and characterize PHOs, ultimately contributing to the protection of our planet.

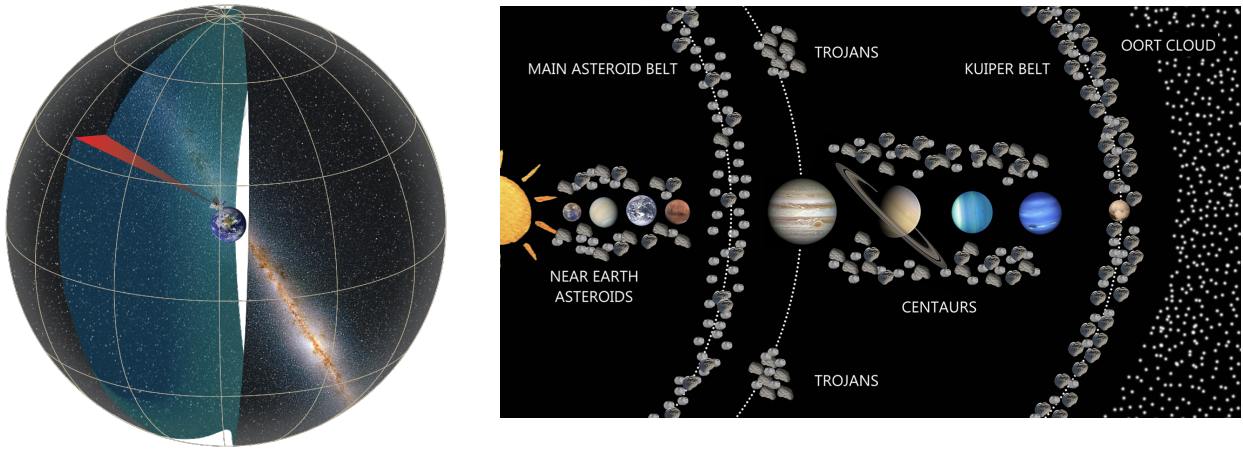

**Figure 1:** (*left*) **SPHEREx all-sky near-infrared mapping.** Utilizing a sun-synchronous NEOWISE-like polar orbit, objects in the sky at ~90 deg elongation will be observed in each great circle. The Earth's motion around the Sun advances the great circle's longitude ~1 deg/day; taking data in both the leading/trailing (forward/ behind) directions means that the entire sky's range of longitudes is covered in 6 months, with the ecliptic poles observed every orbit. (*right*) **The rich set of solar system objects available for SPHEREx study** (except for the Sun, Mercury, Venus, and the innermost NEAs). (After Frantseva 2019.)

## I. Overview of the SPHEREx Mission

### A. Mission Capabilities

SPHEREx is a NASA Astrophysics Division Mid-Sized Explorer (MIDEX) mission designed to create the first survey of its kind, mapping the entire sky in low-resolution near-infrared spectroscopy (Doré+ 2016, 2018). The mission was approved by NASA in February 2019 from a competitive selection, and approved for launch on a Space-X Falcon 9 in early 2025. It is currently in Phase D (integration & testing).

The SPHEREx mission targets 3 core science themes as its baseline science. In the process of conducting these three core NASA Astrophysics investigations (Measuring the Anisotropy of Cosmic Inflation, Determining the History of Galaxy Formation, and Surveying Ices in Molecular Clouds), the mission will collect millions of skyframe exposures in 102 visual and near-infrared wavelength (VISNIR) spectral bands ranging from 0.75 to 5 um, with *each* band at approximately NEOWISE W1 (3.6 um band)/ W2 band (4.5 um band) sensitivity. I.e., SPHEREx will provide 40 discrete spectral channels spanning the previous WISE/NEOWISE survey's 2 photometric channels, **plus** sixty-two spectral channels covering the 3-band JHK 2MASS survey. It will do this using a single instrument system – a passively cooled, high throughput, non-dispersive 6-focal plane patterned LVF system at spectral resolution R = 35 – 130.

Launching in early 2025, SPHEREx, will follow-on for planetary defense (PD) where the highly successful, but retiring-in-2024 NASA PD asset mission, NEOWISE, stops. Unlike the NEOWISE or the NEOSurveyor missions, which obtain critical discovery and thermal-characterization data yielding size estimates with broad band-passes, SPHEREx will sample object spectra at resolutions powerful enough to separate an object's key gas spectral lines and solid state features. Orbiting above the atmosphere, SPHEREx will routinely observe objects in the important $H_2O$, $CO_2$, and CO bands typically inaccessible from groundbased observatories due to the obscuring terrestrial atmosphere.

SPHEREx measurements will be uniquely useful for spectral typing, quick object compositional characterization, population context, size/albedo determination, and temporal trending of objects in the current epoch. PD models that inform mitigation strategies take as key inputs spectral types, sizes, albedos, and rotation states.

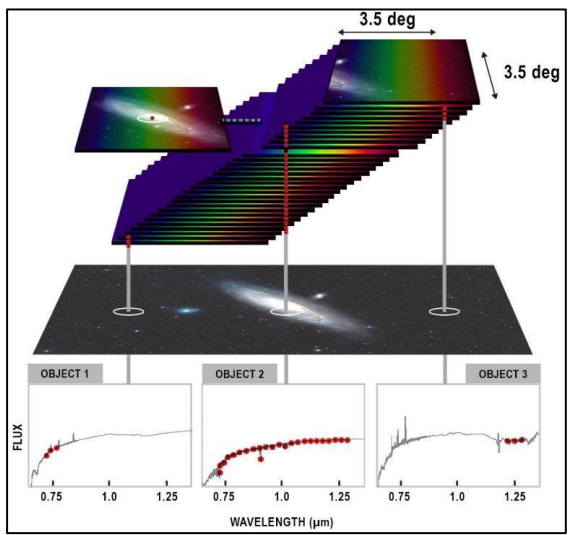

**Figure 2 – Building up an object's complete SPHEREx spectrum requires it to be observed with every separate LVF spectral bin, i.e. placed on 102 different rows of the 6 SPHEREx H2RG focal planes.** Utilizing LVF's rather than dispersive spectroscopic elements ensures SPHEREx's high throughput, stability, and sensitivity at the expense of non-instantaneous spectral data collection.

**B. Surveying the Entire Sky in Near-Infrared Wavelengths**

Utilizing a sun-synchronous NEOWISE-like polar orbit, objects in the sky at ~90 deg elongation (i.e., dawn/dusk) will be observed in each great circle (Fig. 1). The Earth's motion around the Sun advances the great circle's longitude ~1 deg/day; taking data in both the leading/trailing (forward/ behind) directions means that the entire sky's range of longitudes is covered in 6 months with the ecliptic poles observed every orbit. During its two-year prime mission, SPHEREx will map the entire sky 4 times, collecting over 8 trillion spectral samples containing ~$10^5$ foreground Asteroids, KBOs, Trojans, and Comets (Doré+2016, Ivezic+ 2022). A complete object spectrum will be built up over ~2 weeks (Figures 2 and 3).

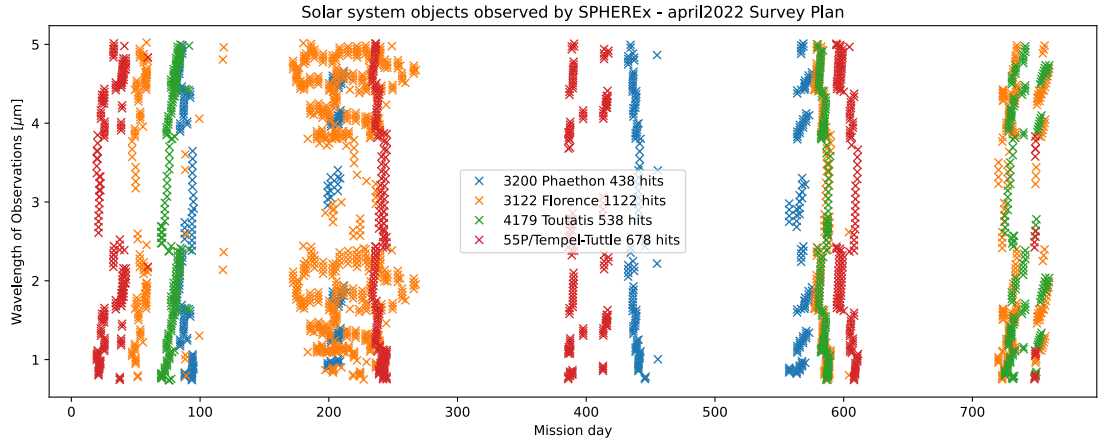

**Figure 3 – (a) Expected SPHEREx LVF wavelength of observation versus time "waterfall" plots** for known asteroidal PHOs 3200 Phaethon, 3122 Florence, 4179 Toutatis, and cometary PHO 55P/Tempel-Tuttle. Each object is acquired at 3 to 5 different epochs over the course of the 2-year prime mission, and observed 100's of times in individual sky images over the course of ~2 weeks in each epoch. It thus takes ~2 weeks to build up a complete SPHEREx 0.7 – 5.0 um spectrum of an object.

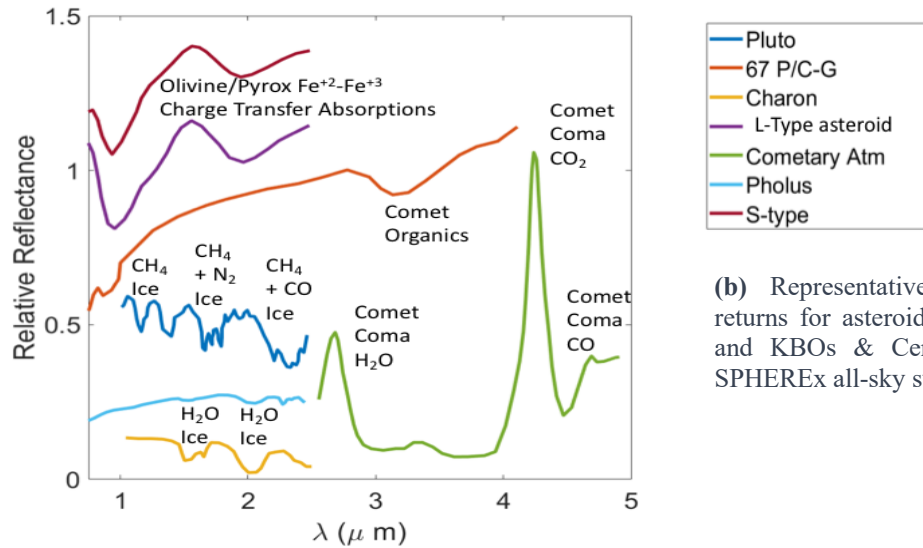

**(b)** Representative expected spectral science returns for asteroids (upper left), comets (right), and KBOs & Centaurs (lower left) from the SPHEREx all-sky survey.

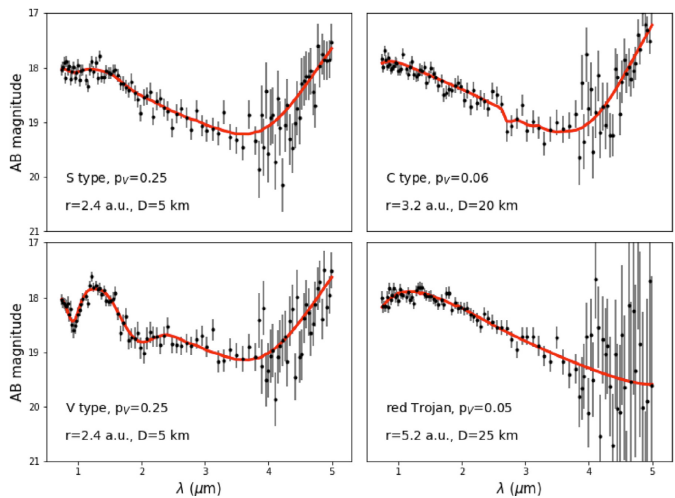

**(c)** Calculated SPHEREx spectra for 3 different kinds of **small** asteroids and Jovian Trojan, including realistic error bars. The taxonomic type, visual albedo $p_v$, heliocentric distance r and object diameter D are listed in each panel. Object sizes are chosen to approximately reproduce a signal-to-noise ratio of 5 at the wavelength of 2.4 μm. Note the small but significant water absorption feature at ~2.7 um for the C-type asteroid, the pronounced 1-2 um silicate absorption features for the V-type asteroid, and that for main-belt asteroids at 2 – 3 au from the Sun, thermal emission dominates past ~4 μm, while it is negligible for Trojans at 5.2 a.u. (After Ivezić *et al.* 2022.)

## C. Data Collection, Processing, and Archiving

The returned SPHEREx data are LVF patterned images of the sky taken every ~2 minutes along great circles on the sky as SPHEREx orbits the Earth every 90 minutes. Each day, SPHEREx will download its sky survey results to the science operations center at Caltech's Infrared Processing and Analysis Center (IPAC) to form a database of all-sky spectral images. Post-processing at IPAC by the SPHEREx science pipeline converts the raw L1 images into calibrated L2 flux images while correcting for dark current, full well/saturation, cosmic ray, and spectral smile effects. A reference catalog of known sky objects is then consulted to find the sources in each one of these L2 images and perform forced photometry on its sky image location, returning a photometric flux measurement for the targeted LVF row's wavelength. Designed to measure the spectra of fixed astronomical sky sources, moving and rotating sources like solar system objects will be detected by this scheme as well - but will require care in stitching together the individual spectrophotometric measurements into a spectral whole due to their time-variable Sun-Earth-Object geometry and projected surface areas.

The steps required to implement this spectral template matching are laid out in a paper written and published by the SPHEREx science team (Ivezic+ 2022, Figure 4):

- Collect all the spectrophotometric data for a given object.
- Obtain ancillary orbital information and correct out any geometric observing factors using a Flux = Flux$_0$ * $(1/r_h)^2$ $(1/\Delta)^2$ $10^{-0.35 * \alpha}$ law (where **α** is the phase angle).
- Correct for systematic flux variations created by rotational modulation of the body's observed surface using template matching (i.e., accurate spectral models of 0.7 – 5.0 um asteroid/Trojan/comet nucleus/KBO spectra) or the body's known lightcurve behavior (e.g., NEOWISE, LSST measurements & PDS data).

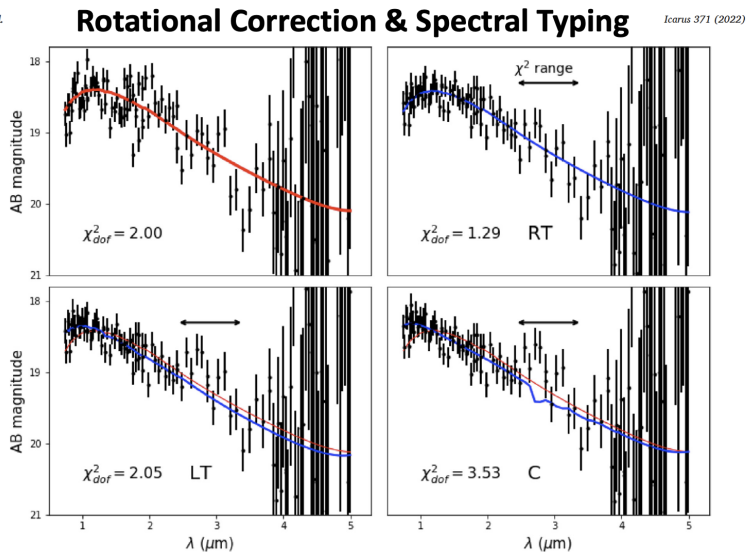

**Figure 4 - Example of SPHEREx spectrophotometric data**, its correction for effects due to an object's motion with respect to the spacecraft over the course of two weeks, and its correction for intrinsic variability (due to projected cross-sectional variations created by a rotating non-spherical body). (***Upper Left***) Simulated "As-Is" data with realistic error bars. The next three figures show the results of geometrically correcting and then forcing the "As-Is" photometry to match different solar system object spectral templates, with the statistically significant lowest chi-squared found for the "red trojan (RT)" template used to generate the data. (After Ivezic *et al.* 2022.)

The resulting spectra will be collated into a searchable database, hereafter termed the SPHEREx Solar System Object Catalog (SSOC), to be submitted to the PDS and IRSA archives for permanent storage. In this SSOC, individual photometric measurements for an object at each wavelength will be collated by time of observation, and the combined, properly corrected 102-band spectrophotometry (Fig. 4) for an epoch of visibility (typically 3 to 5 in total during the 2-year prime SPHEREx mission, Fig. 3) associated with it. Orbital geometrical information and sky image postage stamps of the object at the time of observation will be collated as well. The ~$10^5$ spectra will be orders of magnitude more than are extant today, with none of the degeneracy/uncertainty of broad band photometric measurements.

The gathering and implementation of these steps will be performed by an experienced team based at APL + UMD working in concert with the NASA PD Office. (By experienced we mean a team that has previously worked on the NASA PD related missions Deep Impact, NEOWISE, and NEOSurveyor, as well as the PDS archive small bodies node.) By covering the full sky in the near-infrared, SPHEREx spectra will thus complement concurrent ground-based and space-borne facilities, enabling rich and diverse astronomical investigations.

## II. Use of the SPHEREx Data for Planetary Defense vs. PHOs

### A. The PHO Problem.

PHOs are defined as small bodies on orbits that come within a minimum orbit intersection distance (MOID) of the Earth of 0.05 au or less, and that have an absolute magnitude (H) of 22.0 or less (NEO Sci. Def. Team Report (2017)). (I.e., they are objects that are known to come close enough to the Earth that they might someday impact us, and their size is large enough that the impact could have catastrophic effects on the Earth' surface.) The H magnitude maximum of 22 corresponds roughly to a size of 140m for an asteroid albedo of 0.14, corresponding to the *George E. Brown size limit* mandate and the size regime for a regional impact catastrophe. According to the NEO Sci. Def. Team Report (2017), impacts of this size or larger occur on the order of every few tens of thousands of years. The *George E. Brown congressional mandate* tasks NASA to find 90% of these PHOs, and the contribution of the many current and near-term future surveys, including NEO Surveyor (NEOS) and LSST, will likely achieve this goal within the next ten years.

Accurate spectral categorization of NEOs is a key factor in assessing the threat from a potential impactor as well as developing effective mitigation strategies (cf. Reddy *et al.* 2022 and Mathias *et al.* 2017). Succinctly, whether the impactor is made of rock, metal, or an icy organic mix is critical to know before one attempts to terminate the hazard *("know thy impactor"*), and this determination is typically made using near-infrared spectrophotometry to measure the important 0.8 – 1.2 um silicate absorption bands, the 3 um water absorption feature, and the VISNIR broad band colors (Trilling *et al.* 2016; Reddy *et al.* 2012, 2016, 2018). Thus the SPHEREx spectral survey data will have immediate and important down-home practical applications for the National Preparedness Strategy for NEO Hazards & Planetary Defense (OSTP 2023). Enhancement of the NEO characterization capabilities also falls under the first goal (Action 1.2) of the NASA Planetary Defense Strategy & Action Plan.

### B. Utilizing SPHEREx's Spectrophotometry for PHO Characterization.

With a limiting detection magnitude of K ~ 19, SPHEREx will be not the most powerful infrared *object discovery* asset for PD in the next decade (this will be NEOSurveyor and LSST), but it will be one of the premier *object characterization* assets, augmenting the information obtained from the discovery assets. Though SPHEREx is not designed to discover objects nor provide asteroid diameter radiometry-based measurements with an accuracy better than that of NEOSurveyor, SPHEREx will provide key, unique spectral information of objects currently obtained by follow-up observations at telescopes like the Planetary Defense NASA-IRTF 3.3m telescope asset on top of Mauna Kea, HI. (One can analogously think of SPHEREx as *"SpeX/Prism mode in space"*.) Leveraging overlapping observations from all three platforms will thus provide a rich and accurate categorization of the bodies that may pose any future threat.

Complementing the discoveries and size measurements from other surveys, SPHEREx will: **(1)** Provide unique 0.7 – 5.0 um object spectra; **(2)** Improve the orbital solutions of known objects with additional imaging data from a stable, precise space-based observing platform; **(3)** Inform the uncertainty in the object's observed flux variability to better match the margin of error in the cross-section geometry correction (Figure 4); **(4)** Extend the coverage of any activity, mass-loss, or object evolution (e.g., outgassing, fragmentation, dust emission) behavior.

For example, SPHEREx will support *NEOSurveyor* and *LSST* observations of comets by providing more complete dust continuum spectral measurements, as well as augment existing models for the dust outflow rates (Kwon *et al.* 2021). The potential of producing and utilizing **combined** *LSST/Rubin + SPHEREx + NEO Surveyor* 0.4-8.0 μm spectral energy distributions (SEDs) for each of ~$10^5$ asteroids is similarly enormous, both for producing new science (see Appendix A) and quick, accurate compositional determination of any new impact threat (especially of the relatively featureless C, D, P, and M-class asteroids, where accurate continuum spectral slope measurement is critical).

### C. Mining SPHEREx Data for the Detection of New PHOs.

The SPHEREx project will only actively track and report spectrophotometry solar system objects with cataloged orbits known at the time of the prime mission in 2025 - 2027. Combing through the SPHEREx all-sky photometric image datasets is expected to turn up thousands of new discoveries by community researchers. For example, the LSST/Rubin consortium at the University of Washington plans on using the THOR guided tracklet algorithm (Moyens+ 2021) to search through LSST and SPHEREx data to discover thousands of new asteroids. Thus, with guidance to the community on how to optimally utilize the SPHEREx Solar System Object Catalogue (SSOC), SPHEREx is thus expected to expand the total number of known solar system small bodies and PHOs by an estimated 10 – 20% from its prime mission observations. (It is important to note that SPHEREx, without any currently foreseen mission expendables, could easily obtain sky data thru 2035, expanding the pool of new discoveries even further.)

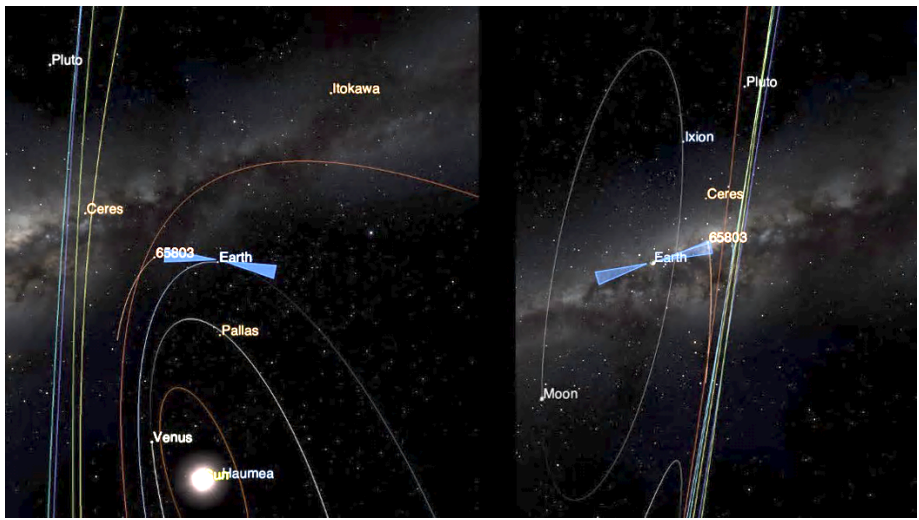

**Figure 5 – A new, dangerous PHO is announced, termed "65803".** Worldwide, observatories will slew to detect it, further measure its apparent sky motion and orbit solutions, and characterize its nature. SPHEREx, with its all-sky survey, will be able to quickly mine its dataset to extract any already extant observations of it in the SSOC, as well as forward-predicting when the s/c will observe it next.

**D. Example Use Case of the SSOC for Planetary Defense**

A potential use case scenario utilizing CNEO SCOUT guidelines (c.f. https:/cneos.jpl.nasa.gov) follows. Assume that an observatory announces the discovery of a new object, which we will call **"65803"** (Fig. 5), that at first orbital solution (derived from small arcs on the sky of the object's apparent motion) has a high probability of hitting the Earth. As telescopes worldwide slew to observe it, its orbit and approximate absolute brightness, H (defined as the apparent magnitude of an object at 1 au from the Sun and Earth and 0 deg phase, roughly proportional to the Albedo*Surface Area product for a fully illuminated disk), become better refined. At the point that these values are stable, the APL + UMD team searches through the SPHEREx catalog using predesigned SSOC data analysis tools, performing a backwards search in time for detections of the object's spectrophotometry and the sky background flux for those detections. Assuming that SSOC spectrophotometry is obtainable, its raw spectrum will be formed and sky images inspected for extension and any co-moving bodies (i.e., fragments or moons). This process will also produce a detrended a set of weeks-long SPHEREx object precovery astrometry at ~3" resolution.

The algorithm of Ivezic+ 2022 will then be used to correct the data vs. other SPHEREx small body spectra, and the result will then be compared to the compendium of thousands of other SPHEREx small body spectra and the object classified as an S-type (rocky), M-type (metallic), or C-type (carbonaceous). Typical albedos for each of these classes will be assumed, and used to determine the objects effective radius & volume; this assumed "quick" albedo value will be compared to self-consistent solutions formed by modeling the 3.5 – 5 um portion of the object's SPHEREx spectrum for thermal emission and fitting the 0.7 – 3.5 um region for light scattering. This process will also produce a detrended lightcurve.

The SSOC database will be continually queried and 65803's photometry refined as orbital solutions change and improve. All results including spectrum, lightcurve, and astrometry will be quickly disseminated to the observer and planetary defense communities to help guide their next observations and hazard mitigation plans. For example, the spectrum of the object will determine whether the PHO is rocky, carbonaceous, or metallic, while the astrometry, reported to the MPC, may improve the first orbit solutions enough to prove that the PHO is not dangerous. The lightcurve data will allow better determination of the object's dimensions, spin state, rough shape, and any mass emission activity and/or mass shedding via fragmentation.

**IV. Summary.**

In the phase of characterizing PHOs for future impact threat mitigation purposes, SPHEREx will support it in many respects. SPHEREx will provide a means of providing unique inputs such as accurate spectral identification and shape information required for Models that provide mass, strength, and compositional constraints needed for mitigation missions or strategies (cf. Mathias *et al.* 2017). The Bus-DeMeo (DeMeo *et al.* 2009) classification system, which extended the previous classification schemes to include the Near-IR, spans the spectral range covered by SPHEREx, but currently only covers a few hundred asteroidal characterizations. SPHEREx will return up to 100,000 spectral characterizations.

By canvassing the entire solar system for 2 years, SPHEREx has the potential to achieve not only a relatively complete sensitivity-limited survey of the solar system's bodies but also the capability to search for spectral variations of these bodies over time. While there is no active mission project planning at this time for SPHEREx activities beyond the prime Feb 2025 – Feb 2027 timeframe, there is no reason at this time that SPHEREx, with no major mission expendables, could obtain additional sky measurements through 2035.

The SPHEREx Solar System object catalog will add a critical component to these capabilities. SPHEREx has complementary advantages over existing space-based telescope platforms like JWST and HST in that the spacecraft obtains relatively unbiased samples of all objects that fall within the survey field of view, yielding spectra that will provide spectral types for the specific objects observed, as well as de-biasable population statistics (cf. Mainzer *et al.* 2014, Bauer *et al.* 2017, Masiero *et al.* 2022) for the sample in total.

For active bodies, the SPHEREx data will provide identification of basic species that drive activities that are difficult or impossible to identify from ground-based observatories (like CO and $CO_2$) and that may provide significant non-gravitational perturbations to the cometary orbits (including those of near-Earth comets (NECs)) over their perihelion.

We end with a quotation: "By harnessing the capabilities of SPHEREx, the scientific community can significantly enhance our ability to detect, track, and characterize near-Earth objects, enabling more effective planetary defense measures. The comprehensive all-sky spectral survey data provided by SPHEREx presents a valuable resource for identifying potential threats and mitigating risks."

**- Appendix A. General R&D Efforts Enabled by the SPHEREx Data Potentially Useful for Planetary Defense. -**

Based on previous mission observational experience, there are numerous scientific findings to be enabled by the SSOC (Doré+ 2018). SPHEREx will spectroscopically survey a broad range of NEOs across their orbital and spectral classes and related populations. These immediate-source populations, such as the Jupiter-family comets and Centaurs, the reservoirs for the NECs and the main-belt asteroids, and the source of origin for the NEAs, will be studied in like fashion. Important science studies we can expect include:

- Determination of the albedo, and *composition* of ~$10^5$ asteroids of different spectral classes (i.e., S vs C/D/P vs V vs M complexes and their sub-classifications within the Bus-DeMeo system (DeMeo *et al.* 2009)), aiding in support of the NASA *Psyche, Lucy,* and other small body missions.
- Discovery of newly active asteroids and characterization of known episodically active asteroids.
- Characterization from 0.7 - 5.0 μm of Interstellar Objects passing through the SPHEREx sky survey from a stable, sensitive, above-the-atmosphere observatory.
- Discovery of 10's, and detection and characterization of 100's of known Centaurs and Comets, leading to better understanding of the origins and evolution of their primordial icy materials (especially $CO_2$ which is unavailable from the ground), as well as the support of the ESA *Comet Interceptor* mission.
- Constraints of size, shape, and rotational properties of NEOs.

**- Appendix B. Speed Limits, Extended Objects, and Coverage. -**

As noted earlier, moving objects will create some challenges in the production of spectra, though we believe our tested procedures will effectively mitigate these complexities. With 6" pixels and a 90 sec exposure time per individual sky frame, an object would have to be moving > 2"/90 sec (.022"/sec, 1.3"/min, or 80"/hr) with respect to the sidereal motion in order to be noticeably smeared in a SPHEREx image. Typical earth reflex motion induced movement for distant solar system objects is typically 1/12th this rate, and even main belt asteroids at rh = 2 au are moving slow enough to not smear. It is only the closest NEAs and inner system small bodies that will have any noticeable smearing.

However, fast-moving objects that significantly trail are likely to be identified but not fully characterized by a forced-photometry pipeline. In post-processing, we plan to do a more complete analysis. These objects, which move in excess of ~1.2 degrees per day, will likely be lower SNR detections but will actually have expanded wavelength coverage if their motion is in the wavelength-dispersion direction.

Comets that are significantly extended may actually provide more complete wavelength samples of coma species and dust. As elements of the comet that fall across different wavelength filter-passes, coma brightness models, with dust and gas components, can be applied, and emission line fluxes can be extrapolated to extract actual species production limits.

### - Appendix C. Short Turnaround Observations for Ultra High Priority Targets. -

A different approach will be necessary for urgently needed short-turnaround results on newly discovered PHAs versus the planned yearly SSOC public delivery. During the prime 2 year 2025-2027 mission, the SPHEREx cosmological sky survey cadence is not planned to be interrupted except for s/c safes or instrument failures, and no pointed observations are planned. There are however means to recover the survey cadence after an interruption, and thus an extremely urgent pointed observation could be entertained in the utmost emergency. For critical but less time intensive measurement requirements, data could be picked out of the survey as soon as it is obtained and rushed through SSSSWG "by hand" processing, with results returned in a matter of weeks. Otherwise objects will be processed in batch, and the results returned in the context of 100s of thousands of other sonar system objects on yearly timescales.

### - Appendix D. Challenges and Considerations. -

#### I. Handling the Large Amount of SSOC Data Generated by SPHEREx will Entail:

- Accurate tracking of millions of observations of moving solar system objects with cross-matching to positions in the SPHEREx sky images at the time of each exposure.
- Development of special tools over and above the normal SPHEREx Science pipeline (see II. below)
- Understanding and cataloging the noise and systematics of the moving object photometry.
- Removing the effects of observational geometry in a uniform way.
- Removing the effects of object rotational variability in a uniform way.
- Putting the individual photometric measurements together to form well understood spectra.
- Pulling spectrophotometry for a few high valued targets, as directed by NASA HQ, quickly from the SPHEREx pipeline.
- Putting the SPHEREx data into PDS format.
- Vetting the PDS data.
- Balancing the budgetary requirements of SPHEREx and planetary defense utilizing a small dedicated team of analyzers and archivers at APL + UMD.
- Delivering the SSOC to publicly usable PDS and IRSA data archives.

## II. Developing efficient algorithms for NEO Detection and Characterization.

Dealing with moving, variable, and oft-times extended objects, the SSOC will require development of special tools over and above the normal SPHEREx Science pipeline, which focuses on distant, fixed sky objects like stars and galaxies. These tools will provide the community user with high-level capabilities to quickly search through the SSOC database and analyze any available NEO data.

These required Tools will include a:

- **Sky Tiler** (stacks multiple visits to a given piece of sky)
- **Postage Stamper** (cuts out & displays a patch of sky at a set time)
- **Background/Shadow Finder** (finds all pointings on a given [RA, Dec]) and displays them
- **Great Circle Imager**: shows SPHEREx's sky-swath for a given UT
- **Lightcurve Extractor** (corrects all points in a given wavelength range for observing geometry; compares to ancillary lightcurve info)
- **SED Extractor** (combines all 102 flux measurements after geometry correction into 1 SED; overlays selected spectral templates; labels when each point was taken; compares to ancillary SED information for an object)